\documentclass[a4paper,fleqn,usenatbib]{mnras}

\usepackage[T1]{fontenc}
\usepackage{ae,aecompl}

\usepackage{graphicx}	
\usepackage{amsmath}	
\usepackage{amssymb}	
\usepackage{pdflscape}	
\usepackage{verbatim}

\usepackage{color}
\usepackage{pgfplotstable}
\usepackage{booktabs}
\usepackage{rotating}

\title[Spectroscopic LF of Abell 85]{Deep spectroscopy of nearby galaxy clusters -- I. Spectroscopic luminosity function of Abell 85}

\author[I. Agulli et al.]{I. Agulli,$^{1,2,3,4}$ \thanks{E-mail: ireagu@iac.es} J. A. L. Aguerri,$^{1,2}$ R. S\'anchez-Janssen,$^{5}$ C. Dalla Vecchia,$^{1,2}$  \newauthor
A. Diaferio$^{3,4}$ R. Barrena,$^{1,2}$ L. Dominguez Palmero,$^{6,1}$ H. Yu$^{7}$\\
$^1$ Instituto de Astrofisica de Canarias, C/ Via Lactea s/n, E-38205 La Laguna, Tenerife, Spain \\
$^2$ Departamento de Astrofisica, Universidad de La Laguna, E-38206 La Laguna, Tenerife, Spain\\
$^3$ Dipartimento di Fisica, Universit\`a di Torino, Via P. Giuria 1, I-10125, Torino, Italy\\
$^4$ Istituto Nazionale di Fisica Nucleare (INFN), sezione di Torino, Via P. Giuria 1, I-10125, Torino, Italy\\
$^5$ NRC Herzberg Institute of Astrophysics, 5071 West Saanich Road, Victoria, BC V9E2E7, Canada\\
$^6$ Isaac Newton Group of Telescopes, Apartado 321, E-38700 Santa Cruz de La Palma, Canary Islands, Spain\\
$^7$ Department of Astronomy, Beijing Normal University, 100875, Beijing, China\\
}

\date{Accepted 2016 February 19. Received 2016 February 19; in original form 2015 December 18}

\pubyear{2015}

\begin{document}
\label{firstpage}
\pagerange{\pageref{firstpage}--\pageref{lastpage}}
\maketitle

\begin{abstract}
We present a new deep spectroscopic catalogue for Abell 85, within 3.0 $\times$ 2.6 Mpc$^2$ and down to $M_{r} \sim M_{r}^* +6$. Using the Visible Multi-Object Spectrograph at the Very Large Telescope and the AutoFiber 2 at the William Herschel Telescope, we obtained almost 1430 new redshifts for galaxies with  $m_r \leq 21$ mag and $\langle \mu_{e,r} \rangle \leq 24$ mag arcsec$^{-2}$. These redshifts, together with SDSS-DR6 and NED spectroscopic information, result in 460 confirmed cluster members. This data set allows the study of the luminosity function (LF) of the cluster galaxies covering three orders of magnitudes in luminosities. The total and radial LFs are best modelled by a double Schechter function. The normalized LFs show that their bright ($M_{r} \leq -21.5$) and faint ($M_{r}\geq -18.0$) ends are independent of clustercentric distance and similar to the field LFs unlike the intermediate luminosity range ($-21.5 \leq M_{r} \leq -18.0$). Similar results are found for the LFs of the dominant types of galaxies: red, passive, virialized and early-infall members. On the contrary, the LFs of blue, star-forming, non-virialized and recent-infall galaxies are well described by a single Schechter function. These populations contribute to a small fraction of the galaxy density in the innermost cluster region. However, in the outskirts of the cluster, they  have similar densities to red, passive, virialized and early-infall members at the LF faint end. These results confirm a clear dependence of the colour and star formation of Abell 85 members in the cluster centric distance.
\end{abstract}




\section{Introduction}
Dwarf galaxies (with absolute bolometric magnitudes, $M_\text{b} > -18$) are the most abundant in the Universe, and their properties differ significantly depending on the local density: unlike the field, where most of dwarfs are star forming, the red, passive population dominates in the clusters, suggesting that the environment preferentially acts in quenching the star formation. The general consensus is that it is a complex mixture of both \textit{mass quenching} -- quenching processes internal to the galaxy -- and \textit{environmental quenching} -- externally driven quenching processes -- that drive the evolution. Multiple studies suggest that different processes dominate the quenching at distinct stellar masses, and that there is a characteristic stellar mass scale between the two types of quenching, with the former dominating in bright galaxy transformation and the latter in the low-mass galaxy evolution \citep[e.g.]{peng2010,wetzel2013,peng2015,weisz2015,davies2016}. However, the \textit{environmental quenching} of dwarfs is far from ubiquitous \citep[e.g.][]{phillips2014,phillips2015,wheeler2014}. 

The relevant physical mechanisms that can be considered \textit{environmental quenching} are the loss of galaxy gas reservoirs \citep{quilis2000,bekki2002}, the gas cooling time \citep{wr1978} and the suppression of dwarfs due to a combination of feedback, photoionization or/and dynamical processes \citep[e.g.][]{moore1996,benson2002,benson2003,aguerri2009,brooks2013}. However, the interplay of the different physical processes and their time-scales are not clear \citep[e.g.][]{balhog2004,baldry2006,peng2010,wijesinghe2012,wetzel2013,fillingham2015}.

Deep spectroscopic surveys to study dwarf galaxies in high-density environments of different physical properties are needed to constrain the physical mechanisms influencing dwarf evolution and to better understand the environmental effects on their evolution \citep[e.g.,][]{sanchez2008}. We have undertaken a campaign of deep spectroscopic observations of nearby clusters down to the dwarf regime ($M_r \gtrsim M_r^*+4$) in order to infer accurate cluster membership and study the properties of low-mass galaxies, such as star formation histories and orbital parameters. The richness class 1 cluster Abell 85 (A\,85 hereafter) is a nearby ($z = 0.055$) and massive cluster \citep[$R_{200} = 1.02 \, h^{-1} \,$Mpc and $M_{200} = 2.5 \times 10^{14} \; M\odot$][]{rines2006}. It is one of the brightest clusters in the X-ray sky \citep{edge1990} with a smoothed gas distribution. It has been extensively studied in the whole electromagnetic spectrum: in the radio \citep{slee2001}, in the optical \citep{rines2006,ramella2007,bravo2009} and in the X-ray \citep[e.g.][]{schenck2014,ichinohe2015}. The main cluster hosts an X-ray bright, metal-rich cool core and it is well known that A\,85 is not completely virialized. Its substructure was analysed in the literature at different wavelengths \citep[see][]{ramella2007, boue2008, aguerri2010}. \cite{schenck2014} found that A\,85 shows a complex X-ray temperature structure, and \cite{durret2003} detected an extended X-ray peak in the south-west of the cluster. Moreover, other two Abell clusters, A\,87 and A\,89, are close in projection to A\,85 \citep{abell1989}. The former is also the nearest and is more likely to be a series of groups moving towards the main body of A\,85 \citep{durret1998}. Both the X-ray emission and the optical galaxy distribution trace an elongation of the cluster orientation towards the A\,87 direction. Therefore, A\,85 is a perfect starting point for our project due to its physical characteristics together with a few practical reasons: its sky region is covered by the Sloan Digital Sky Survey (SDSS), so bright galaxies ($m_r \leq -18$) already have spectroscopic information, allowing us to focus on the faint ones, and its proximity favours deep spectroscopic and large area coverage at the same time. 

The galaxy luminosity function (hereafter LF) gives the number density of galaxies of a given luminosity. It is a powerful observable to study the characteristics of galaxies and to compare environments of different densities \citep[e.g.,][and references therein]{trentham2002,trentham2005,tully2002,blanton2003,infante2003}. In particular, the faint end seems to be shaped by the environment \citep[e.g.,][]{popesso2006,barkhouse2007}, making deep spectroscopic study of dwarf galaxies and the comparison between those in field and clusters fundamental. The Schechter function \citep{schechter1976} is commonly used to fit the LF of galaxies. This function is exponential at the bright end and a power law at the faint end. However, its universality is still a matter of debate. While it is a good fit to some (spectroscopic) observations of clusters \citep[e.g.,][]{rines2008}, a linear combination of two Schechter functions is needed in other cases \citep[e.g.,][]{blanton2005,popesso2006,barkhouse2007}. 

Previous studies suggested that the most remarkable difference between the global and cluster LFs is in the faint-end slope \citep[e.g.][and references therein]{blanton2005, popesso2006}. Photometric studies show that the LF in clusters has a steeper faint end. However, this is not the case for some nearby galaxy clusters with deep spectroscopic LFs \citep[e.g.][]{rines2006}. \cite{agulli2014} showed that the LF of A\,85 measured in three decades in luminosities has an upturn at $M_r \sim -18$ and a steep faint end. Although the faint end is less steep compared to the photometric results of \cite{popesso2006}. A\,85 LF shares common features with the spectroscopic field LF by \cite{blanton2005}, in particular a similar faint-end slope. The main difference between the A\,85 and the field LF is the nature of the low-mass galaxies. Thus, red dwarfs dominate the cluster faint end of the LF while blue faint galaxies are more abundant in the field. These results suggest that the environment plays a major role just in their evolution, transforming field blue dwarfs into cluster red ones, but not in the determination of the LF faint-end slope -- i.e. in the relative abundances of low-mass galaxies. 

The aim of this paper is to study the upturn of the LF of A\,85: its dependence on clustercentric distance, on galaxy type, and on the dominant physical processes that shape its profile.

The paper is organized as follows: the data set is presented in Section 2. In Section 3, we present the classification of the members in different populations. Section 4 describes and analyses the LF of the cluster, its radial dependence and the influence of the different types of galaxies on the upturn. The results are discussed in Section 5 and the conclusions are presented in Section 6. Throughout this work, we have used the cosmological parameters $H_0 = 75 \; \mathrm{km} \, \mathrm{s}^{-1} \mathrm{Mpc}^{-1}$, $\Omega _\text{m} = 0.3$ and $\Omega _{\Lambda} = 0.7$.

\section{The observational data on A\,85}
\subsection{Target selection and spectroscopic observations}
\begin{figure}
\centering
  \includegraphics[width=1\linewidth]{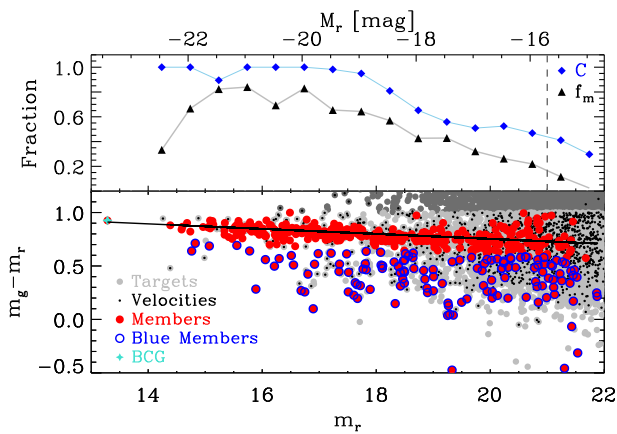}
  \caption{Lower panel: CMD of the galaxies in the direction of A\,85. Grey dots are the target galaxies and black points are the velocities obtained. Red and blue symbols show red and blue cluster members, respectively. The solid line represents the red sequence of the cluster. Upper panel: spectroscopic completeness  ($C$, blue diamonds) and cluster member fraction ($f_\text{m}$, black triangles) as a function of $r$-band magnitude. The dashed vertical line represents our limit magnitude for the spectroscopic LF.}
\label{cmd}
\end{figure}

We carried out an extensive spectroscopic study of the galaxies along the A\,85 line of sight. Our parent photometric catalogue contains all galaxies from the Soan Digital Sky Survey Data Release 6 (SDSS-DR6) \citep[][]{adelman2008}, brighter than $m_r = 22$ mag\footnote{The apparent magnitudes used are the dered SDSS-DR6 $r$-band magnitudes. Note that these magnitudes have been corrected for extinction.} and within 3.6 $\times$ 3.6 Mpc$^2$. From this catalogue, we selected those galaxies with no measured spectroscopic redshifts and with $g-r$ colour bluer than $1.0$ (see Fig. \ref{cmd}) identifying 3082 targets. $g-r = 1.0$ is the typical colour of a 12 Gyr old stellar population with [Fe/H] = +0.25 supersolar metallicity \citep{worthey1994}. The fraction of cluster members lost due to this colour selection is expected to be very small, and we match at the same time the colour distribution of the nearby Universe \citep[see][]{hogg2004,rines2008}.
The brightest cluster galaxy (BCG, $\alpha$(J2000): $00^h \, 41^m \, 50.448^s$ $\delta$(J2000): $-9^{\circ} \, 18'  \, 11.45''$) is assumed to lie at the centre of the cluster. This is a sensible choice given that the peak of the X-ray emission is offset only 7 kpc from the BCG centre \citep{popesso2004}.  Figure \ref{cmd} shows the colour--magnitude diagram (CMD) of the full sample, with the colour cut applied. 

The majority of the observations were performed with the VIsible Multi-Object Spectrograph at the Very Large Telescope (VIMOS@VLT), in combination with the LR-blue+OS-blue grisms and filters (Program 083.A-0962(B), PI R. S\'anchez-Janssen, 2009 August). To maximize the number of targets, and avoid the gaps between the instrument CCDs, we designed 25 masks with large overlaps covering an area of 3.0 $\times$ 2.6 Mpc$^2$ around the central galaxy of A\,85 -- i.e. extending out to more than 1\,$R_{200}$. This strategy allowed us to obtain 2861 low-resolution spectra (R=180). Integrations of 1000 s were sufficient to obtain signal-to-noise ($S/N$) ratios in the range 6--10 down to the limiting magnitude. 

Three more pointings have been added using AutoFib2/WYFFOS, a fibre spectrograph at the William Herschel Telescope (AF2@WHT, service program SW2013b24,  2014 September) covering the central area of the cluster. These observations gave us 176 low-resolution spectra (R=280 and grism R158B) reaching $m_r = 19.5$. The data were collected during three exposures of 1800\,s per pointing in order to reach $S/N > 5$. 

We used the provided pipelines of the two instruments (version 2.25 for AF2 data) to perform the standard data reduction: bias, flat-field, sky subtraction and wavelength calibration \citep{izzo2004,dominguez2014}. For the latter calibration we used an HeNe lamp, which yield a wavelength accuracy  of $\sim 0.5 \, \rm \AA$ pixel$^{-1}$ in the full spectral range ($3700 - 6700 \, \rm \AA$), for the VIMOS data, and an He lamp for the WYFFOS spectra.

\subsection{Spectroscopic galaxy catalogue}\label{tg}
We used the \textit{rvsao.xcsao} \textsc{iraf} task \citep{kurtz1992} to determine the redshifts of the observed galaxies. This task cross-correlates a template spectrum library \citep[in this work,][]{kennicutt1992} with the observed galaxy spectrum and applies the technique described in \cite{tonry1979}. We determine 2114 velocities between the two observing runs. The remaining spectra had too low $S/N$ to estimate reliable redshifts. Formal errors of this task are smaller than true intrinsic ones \citep[e.g.][]{bardelli1994}. This is because the error on the redshift estimate depends on several factors. First of all, the resolution of the grating used in the observations, which, in this case, gives a sigma around 650--700 km\,s$^{-1}$ at 5000 $\AA$ with slits of 1 arcsec. Note that this is the expected value for the resolution of VIMOS, which is worse than the resolution of AF2. Other sources of uncertainty are the wavelength calibration, errors in the cross-correlation and errors in the centring of the targets. The convolution of these can be evaluated only by comparison of different measurements of the same objects. Thanks to our observational strategy we obtained 676 repeated spectra with VIMOS and nine with WHT pointings, resulting in a $1\sigma$ velocity uncertainty of $\sim 500$ km s$^{-1}$ and  $\sim 200$ km s$^{-1}$, respectively. Although these uncertainties are large, they do not significantly affect the determination of the cluster membership. We found that in the range v$_{\text{c}} \pm 5 \, \sigma_{\text{c}}$, where $v_{\text{c}}= 16681$ is the cluster recessional velocity and $\sigma_{\text{c}}= 979$ its velocity dispersion, only 14 galaxies are discarded by the caustic method (see Section \ref{memb}). Given that the number of non-member galaxies is relatively low for such a large velocity interval, we are confident that the bias in the cluster membership determination is negligible.

Combining our data with redshifts from the SDSS-DR6 and NASA/IPAC Extragaalctic Database (NED) catalogues results in a total number of 1603 redshifts in the magnitude interval $13 \leq m_r \leq 22$, within 1.4\,$R_{200}$ and for galaxies bluer than $g -r  = 1.0$. The completeness of the data is shown in the upper panel of  Fig. \ref{cmd}. It is defined as $C=N_{z}/N_{\text{phot}}$, where $N_{z}$ is the number of measured redshifts and $N_{\text{phot}}$ the number of photometric targets. $C$ is larger than 90\% for galaxies with $m_r < 18$ and decreases around 40\% at $m_r \sim 21$.

\subsection{Cluster membership}\label{memb}
We have used the caustic technique to determine the cluster membership \citep{diaferio1997,diaferio1999,serra2011}. When plotted on the plane of the line-of-sight velocity -- projected clustercentric distance, the cluster galaxies occupy a specific and well-defined region with a trumpet shape: a decreasing amplitude, $A$, with increasing clustercentric distance, $r$. \cite{diaferio1997} shoe that $A(r)$ is a function of the cluster escape velocity modulated by a function depending on the galaxy orbit anisotropy parameter, $\beta$.  One of the advantages of the caustic technique is that it measures the escape velocity and the mass profile of the cluster in both its virial and infall regions, where the assumption of dynamical equilibrium does not necessarily hold. As a byproducts, tho technique identifies the cluster members, namely those galaxies whose radial velocity is smaller than the escape velocity at a given clustercentric radius. Numerical simulations show that the cluster membership obtained using this technique has an interloper contamination of only 2\% within $R_{200}$ \citep{serra2013}. 

\begin{figure}
\centering
  \includegraphics[width=1\linewidth]{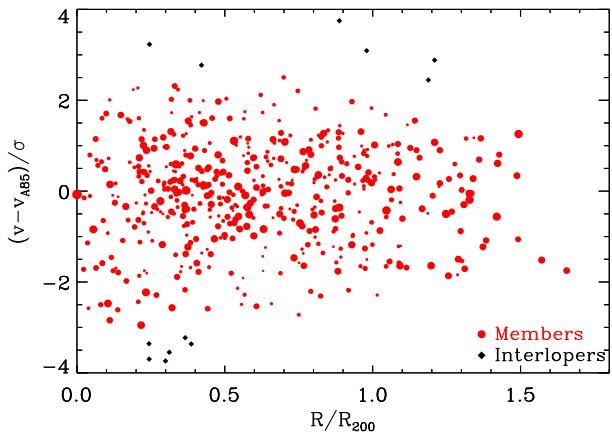}
  \caption{Line-of-sight velocity -- projected clustercentric distance plane for the A\,85 members (red dots) and the interlopers (black diamonds). The red dot dimensions scale with the magnitude: bright (faint) -- big (small) dots.}
\label{caust_memb}
\end{figure}

In order to estimate $A(r)$ for A\,85, we used our spectroscopic data set within the VIMOS field of view, completed with SDSS-DR6 spectroscopic velocities up to 1.7\,$R_{200}$. We adopted this strategy to better constrain the trumpet shape of the caustics and consequently the profiles of the escape velocity and the mass of the cluster. We obtained a sample of $497$ cluster members within 1.7 $R_{200}$; for further analysis, we use only the $460$ members that are within the observed area. The members of A\,85 are shown in Fig. \ref{caust_memb} where they are plotted on the line-of-sight velocity -- projected clustercentric distance plane.

The member fraction defined as $f_\text{m} = N_\text{m}/N_z$, where $N_\text{m}$ is the number of members, is plotted in the upper panel of Fig. \ref{cmd}. It is a function of galaxy luminosity and is larger than 60 $\%$ for $M_r < -19$ and then rapidly decreases to $\sim$ 20 $\%$ at $M_r = -16$.

\begin{figure}
\centering
 \includegraphics[width=1\linewidth]{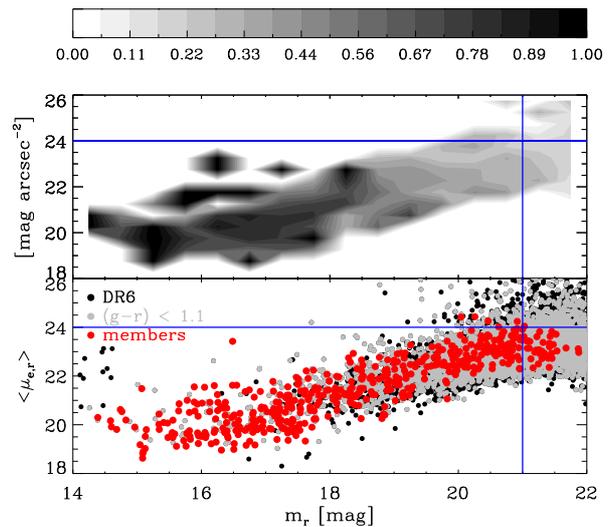}
 \caption{The upper panel shows the completeness as a function of both the apparent magnitude and the mean surface brightness using grey-scale colour levels. In the lower panel the mean surface brightness as a function of apparent magnitude is plotted: the whole SDSS-DR6 catalogue with black dots, the targets with grey dots and the members with red dots. The vertical and horizontal blue lines are the catalogue thresholds for the completeness.}
\label{binned_compl}
\end{figure}

The completeness as function of both apparent magnitude and mean surface brightness is presented in Fig. \ref{binned_compl}. It shows that the target population is well represented by the galaxies with spectroscopic information up to $m_r = 21$ mag and $\langle \mu_{e,r} \rangle \leq 24$ mag arcsec$^{-2}$.

\section{Galaxy classification}\label{gc}
In this work, we aim to understand the type of galaxies producing the main features of the LF of A\,85 and their location in the cluster. Therefore, we separated the cluster members into different populations according to their colour, their dynamics and their star formation. In the following sections we give the details of our classifications.

\begin{figure*}
\centering
 \includegraphics[width=1\linewidth]{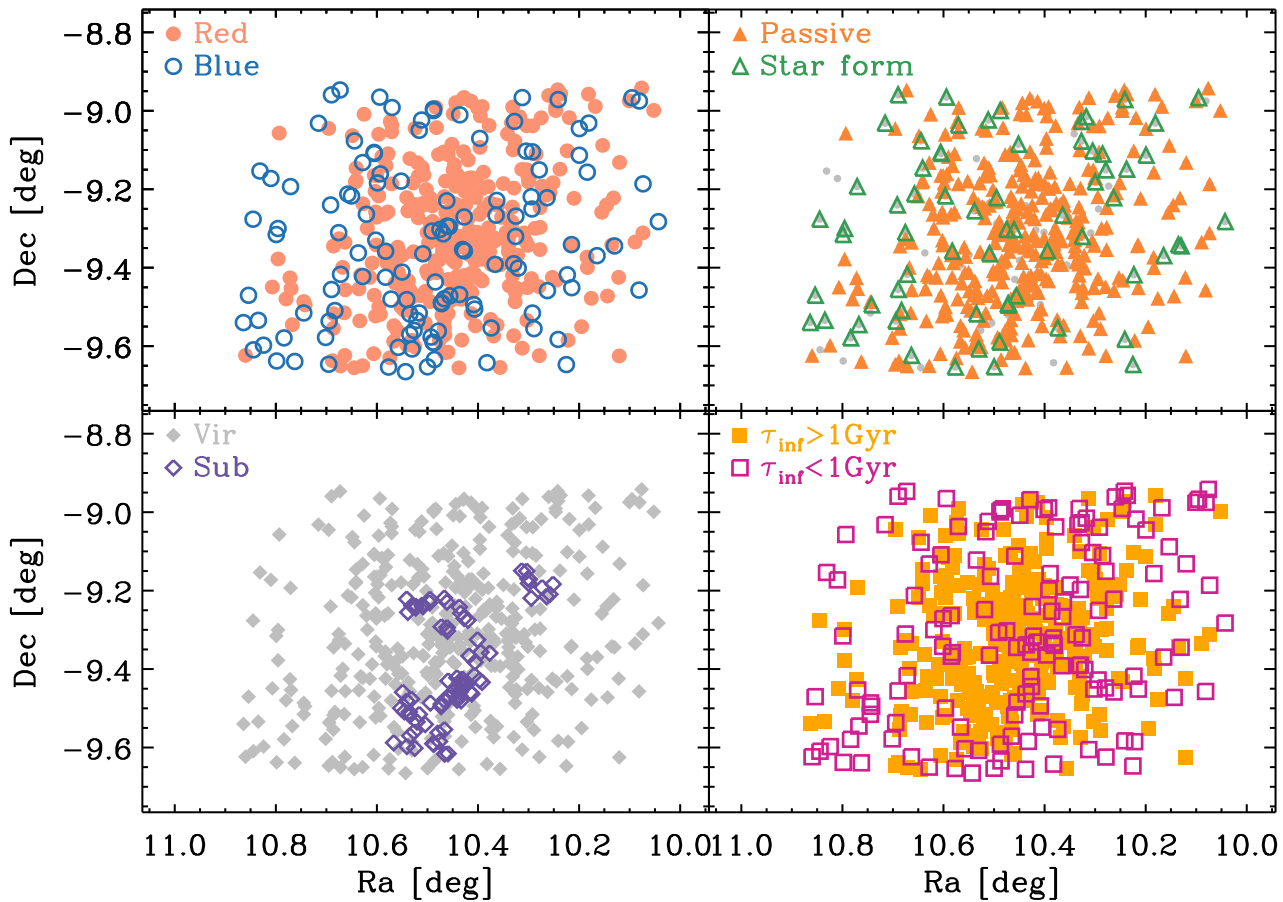}
 \caption{The plots show the positions of the different populations. The upper-left plot presents the red (red dots) and blue (blue circles) members, the upper right the passive (open orange triangles) and star-forming (empty green triangles) populations (due to the $O\,\textsc{ii}$ emission line, in grey dots the members that cannot be classified), the lower-left shows the members in substructure regions (open purple rhombi) or not (filled grey rhombi) and the lower right the members older (filled orange squares) or younger (open magenta squares) that 1 Gyr.}
\label{ra_dec_pop}
\end{figure*}

\begin{figure*}
\centering
 \includegraphics[width=1\linewidth]{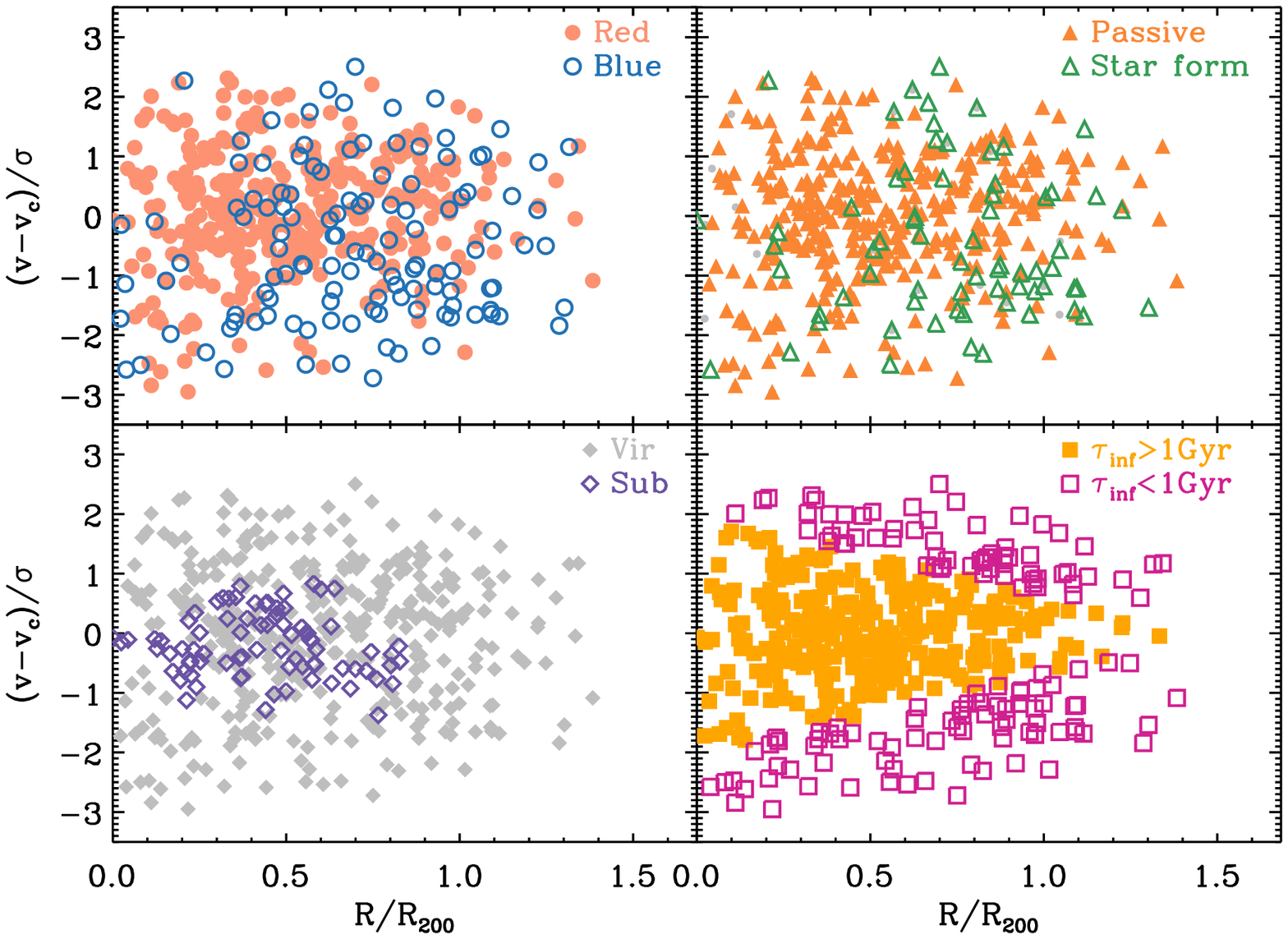}
 \caption{The phase-space of the different populations is shown with the same distributions and symbol and colour code as Fig. \ref{ra_dec_pop}}
\label{caustic_pop}
\end{figure*}

\subsection{Red and blue members}
We separated the members into red and blue populations by applying a colour cut. We define as blue the galaxies bluer than $(g - r)_\text{RS} -3\sigma_\text{RS}$, where $(g - r)_{RS} = -0.02\,m_r+1.22$ and $\sigma_\text{RS} = 0.05$ are the colour and the dispersion of the red sequence of the cluster. The red population is formed by the remaining cluster members. We estimate the red sequence by using the members with apparent $r-$band magnitude brighter than 19 and $g-r$ colour redder than 0.6. This strategy was adopted in order to have smaller uncertainties on the red sequence and a better determination of the two populations. Indeed, SDSS-DR6 errors on the magnitude increase with their values and are larger for low-mass galaxies. The resulting fractions of red and blue galaxies turned out to be 74\% and 26\%, respectively. 

\subsection{Star-forming and passive galaxies}
By using the equivalent width (EW) of the $[O\,\textsc{ii}]$ emission line, we have separated the galaxies into passive and star forming. We measured the EW($[O\,\textsc{ii}]$) using VIMOS and SDSS-DR6 spectra. No EW($[O\,\textsc{ii}]$) measurements could be obtained for those galaxies with only NED or AF2@WHT velocity information. This reduced the number of cluster members by 10\% for this classification. 

Galaxies with EW($[O\,\textsc{ii}]$) $< -5 \, \AA$ are classified as star forming. This upper limit was confirmed by visual inspection of the spectra. Only 16\% of the analysed cluster members turned out to be active star-forming galaxies, as expected for nearby clusters \citep[see][]{pracy2005}.

\subsection{Virialized and non-virialized members}
Our catalogue contains the largest number of spectroscopically confirmed members of A\,85 presented so far in the literature, including 234 dwarf galaxies that were not taken into account in previous studies. Several works showed that the results on the substructure analysis in clusters strongly depends on the population of galaxies used \citep[e.g.,][]{aguerri2010}. To avoid biases, we ran several statistical tests in order to detect substructure in A\,85. We used tests by taking into account only the recessional velocity of the galaxies \citep[1D tests, eg. Kolmogorov-Smirnof and 1D DEDICA][]{pisani1993}, their positions on the sky \citep[2D tests, eg.][]{pisani1993}, or both their positions and velocities \citep[3D tests, eg.][]{ds1988}. 1D tests returned no significant substructures, while 2D and 3D tests gave evidence for substructure.

In order to confirm which galaxies belong to substructures, we applied the caustic method as described in \cite{yu2015}. This algorithm gave 18$\%$ of A\,85 members in substructures.  For a more detailed analysis of A\,85 substructures, their properties and their comparison with X-ray observations see Yu et al. (in preparation) 

\subsection{Early and recent infall members}
According to \cite{oman2013}, the members of a cluster can be classified depending on their infall time. Indeed, the phase space can be divided using the line: 
$$
| \frac{V-V_\text{c}}{\sigma} |= -\frac{4}{3}\frac{\mathrm{R}}{\mathrm{R}_\text{vir}}+2 , 
$$
 where $V$ and $V_c$ are the recessional velocities of the galaxies and of the cluster, respectively, $\sigma$ is the cluster velocity dispersion, R is the projected clustercentric distance and R$_\text{vir}$ is the cluster virial radius. Their simulations show that at a given distance from the cluster centre, a significant fraction of the galaxies above this line fell into the cluster during the last 1Gyr. Applying this analysis to A\,85, we found that 31$\%$ of the members have been recently accreted. 

\subsection{Spatial and phase-space distributions of the different galaxy types}
\begin{figure*}
\centering
 \includegraphics[width=1\linewidth]{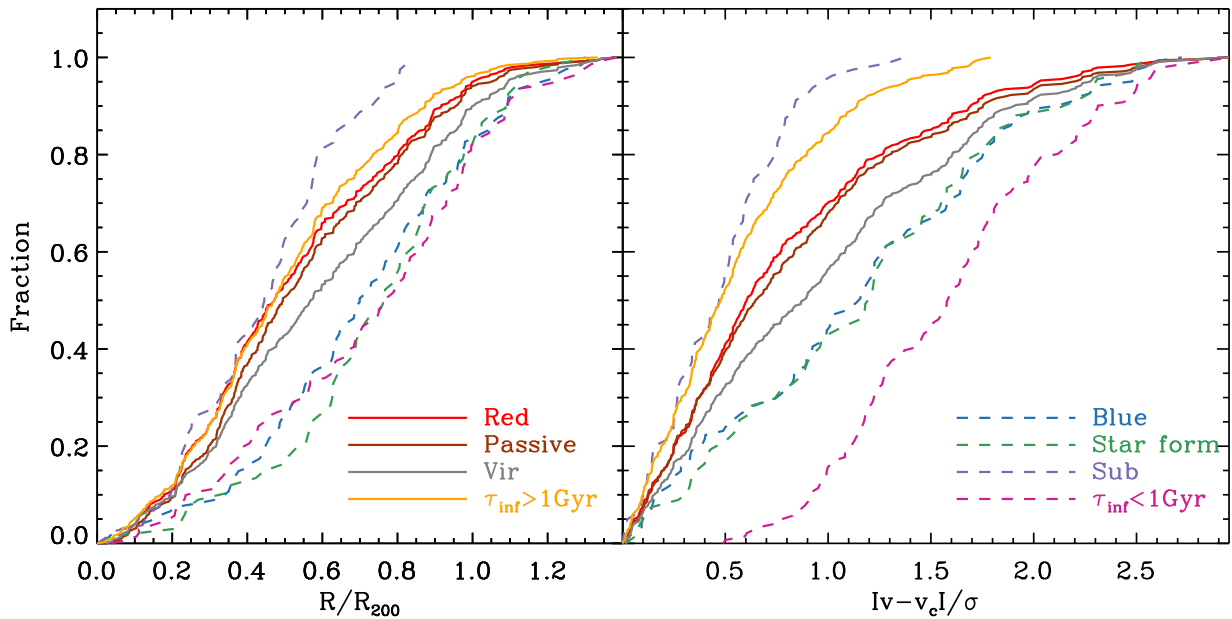}
 \caption{The cumulative distribution function for the different types of galaxies as a function of the radius in the left-hand panel and of the recessional velocity in the right one. The colour code is the same as Fig. \ref{ra_dec_pop}, with the dominant populations in solid lines and the others in dashed lines.}
\label{acum_distr_func}
\end{figure*}

Figures \ref{ra_dec_pop} and \ref{caustic_pop} show the spatial and phase-space distribution of the different types of galaxies according to our classification. Fig. \ref{acum_distr_func} shows the cumulative distribution function of R$/R_{200}$ and $|v-v_\text{c}|/\sigma$ for the same subsets. Red, passive and early-infall galaxies are generally located at smaller clustercentric distances than blue, star-forming and recent accreted members. Moreover, the latter group of populations have higher velocity dispersion when compared to the global cluster velocity dispersion. These trends give a radial and infall-time dependence of the star formation. Galaxies in substructure present a very narrow range in velocities, and are mostly located relatively close to the centre of the cluster. The global properties of the galaxy populations in A\,85 are similar to those observed in other nearby galaxy clusters \citep[see e.g.][]{goto2005, aguerri2007, sanchez2008}.

\section{Spectroscopic LF of A\,85}
The LF is a basic statistics that provides information about the nature of the galaxy populations in different environments. Most of the results on the properties of cluster galaxies have been obtained with the extensive analysis of photometric LFs of nearby clusters. However, few studies have been done based on deep spectroscopic LFs.

\begin{figure*}
\centering
 \includegraphics[width=1\linewidth]{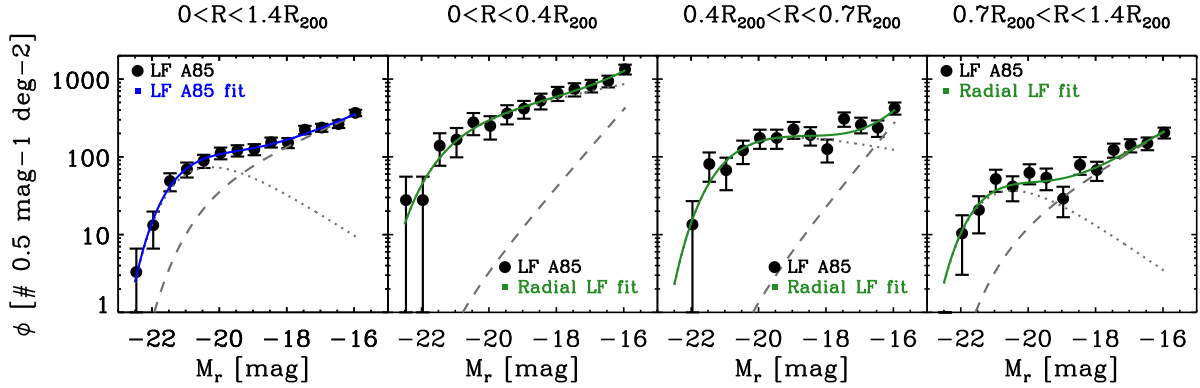}
 \caption{The first left-hand plot shows the total LF of the cluster (black dots) with the double Schechter fit (blue line) and its two components (bright: dotted black line, faint: dashed black line). The other three plots present the radial LF of A\,85 with the best fit as green line and the two components as the first plot. As reported in the upper part of each plot they are, in order, the inner region, out to 0.4\,$_{R200}$, the median one from 0.4 to 0.7\,$R_{200}$ and the outskirts from 0.7 to 1.4\,$R_{200}$.}
\label{lf_rad}
\end{figure*}

We computed the total spectroscopic LF of A\,85 following the prescription in \cite{agulli2014}. The LF is given by: $\phi(M_r) = N_\text{phot}(M_r) \times  f_\text{m}(M_r) / (\Delta M_r \times A)$, where $N_\text{phot}$ is the number of photometric targets (see section \ref{tg}), $f_m$ the fraction of cluster members defined in section \ref{memb}, $A$ the surveyed area and $\Delta m_r = 0.5$ the magnitude bin size used. Fig. \ref{lf_rad} shows the spectroscopic LF of A\,85 and the best-fitting double Schechter function. Indeed, according to the Pearson test, a single Schechter function is not a good representation of the observed LF. In order to reduce the degeneracy of the fit, we used a double Schechter function with only one $M^*$, similarly to previous studies \citep[e.g.,][]{blanton2005}. Therefore, the free parameters of the fit are the characteristic luminosity of the whole cluster, two different slopes for the bright, $\alpha_\text{b}$, and the faint, $\alpha_\text{f}$ components, and their relative normalization. We use this fit function for the whole analysis carried out in this paper. 

The parameters of the global LF are reported in table \ref{tab_rad_fit}. Due to the larger number of galaxies in the sample  and the different double Schechter fit used, they are slightly different from those of \cite{agulli2014}. However, the overall shapes of the fitted function are very similar, including the position of the upturn and the value of the faint-end slope.  

\begin{figure*}
\centering
 \includegraphics[width=1\linewidth]{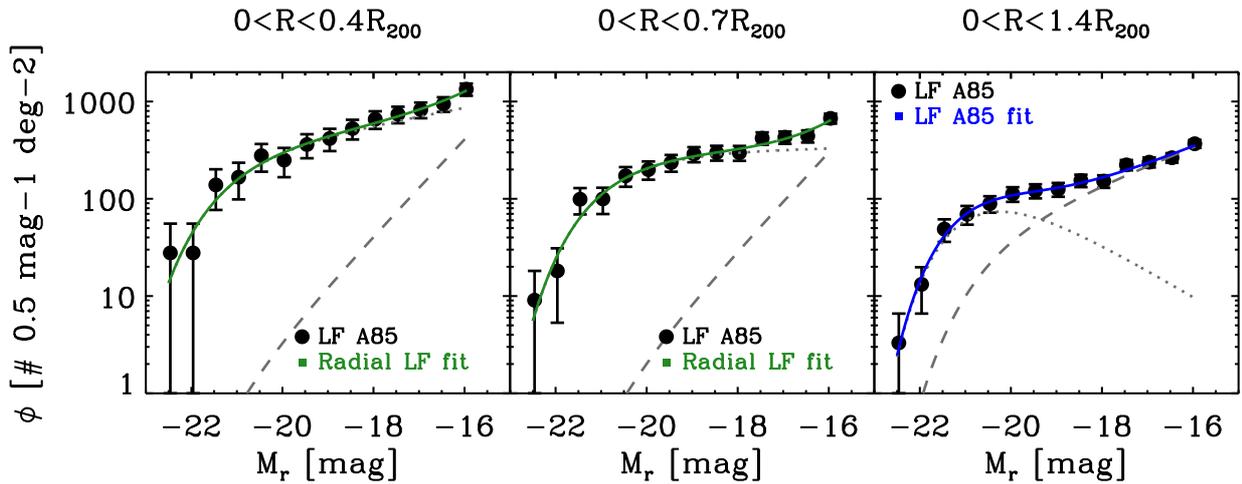}
 \caption{The first left-hand plot is the inner LF, from 0 to $0.4 \, R_{200}$, then the medium aperture, from 0 to $0.7 \, R_{200}$ and in the end the total LF of A\,85, that corresponds to the maximum aperture, from 0 to $1.4 \, R_{200}$. Each of them presents the observed LF in black dots, the double Schechter fit as a solid line and the two components with dotted (bright) and dashed (faint) lines.}
\label{lf_aper}
\end{figure*}

\subsection{The radial LF}
We studied the radial dependence of the LF of A\,85 to analyse the influence of the environment on the galaxy population. We divided the cluster into three regions, an inner circle of radius $r = 0.4 \, R_{200}$, an intermediate ring with $0.4 \, R_{200} < r \leq 0.7 \, R_{200}$ and the outer area with $0.7 \, R_{200} < r \leq 1.4 \, R_{200}$. This choice has no specific physical reason, but each bin contains roughly the same number of members giving similar statistical significance. The resulting LFs are shown in Fig. \ref{lf_rad}. According to the Pearson test, a single Schechter function is not a good description of the data. We have then fitted a double Schechter function (see table \ref{tab_rad_fit}). The relative contribution of the two components of the fit is radial dependent. Indeed, the second component becomes more prominent in the outer LF, and the relative contribution of the first one decreases with clustercentric distance. The former component is the one strongly contributing to the upturn in the external regions of the cluster. A dip is also present highlighting the change in the slope. It is located at different magnitudes, moving from $M_r \sim -18$ in the intermediate LF to $M_r \sim -19$ in the outer radial bin. 

We also studied the influence of different apertures on the LF. In particular, we analysed the cumulative LFs within apertures of 0.4, 0.7 and 1.4 $R_{200}$. The resulting LFs are shown in Fig. \ref{lf_aper}. As in the previous analysis, a double Schechter function is used to properly model the data and the fitted parameters are listed in table \ref{tab_rad_fit}. Here, we do not observe any dip and the second component of the fit dominates only when the galaxies from the most external region of the cluster are considered.

Both the visual inspection of Figs \ref{lf_rad} and \ref{lf_aper} and the analysis of the fitted parameters give a variation of the faint-end slope, $\alpha_\text{f}$, with radius. In particular, the faint end is less steep in the outer region of the cluster. There is also an evolution of $M_r^*$ from the cluster centre to the outskirts: it becomes fainter at larger radii. In the outskirts of the cluster the LF is similar to the total and the field spectroscopic LF \citep[see][]{blanton2005}. 

\begin{figure*}
\centering
\includegraphics[width=1\linewidth]{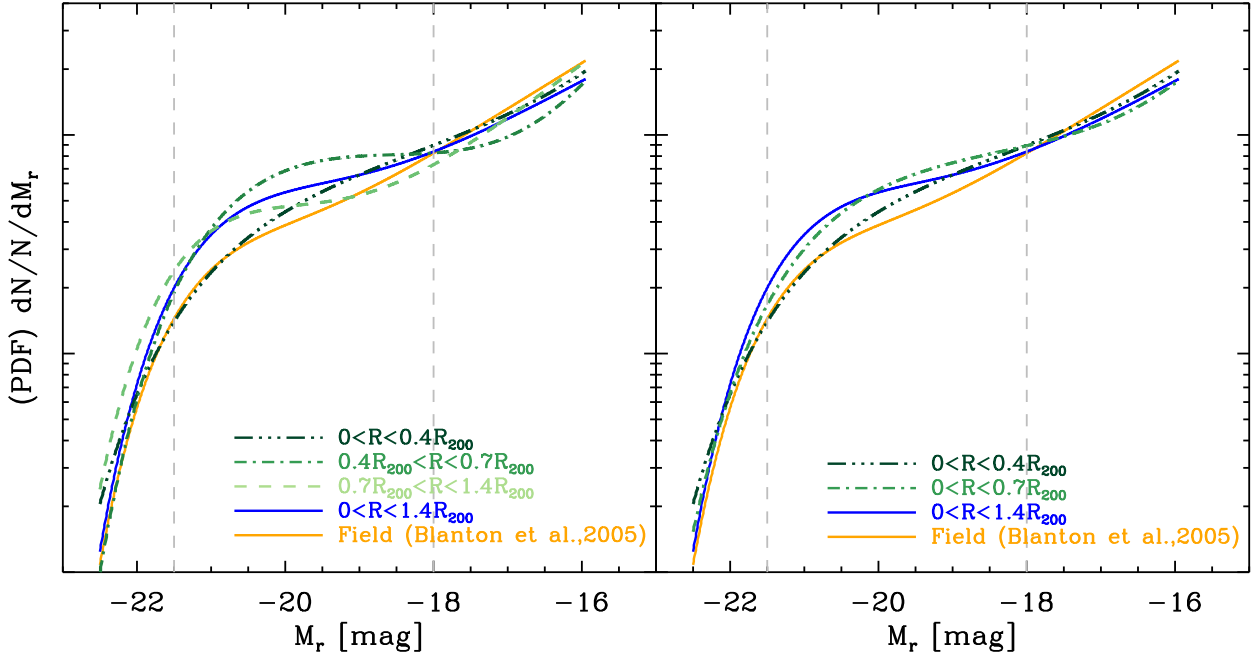}
\caption{The left-hand panel shows the normalized (with the total luminosity) double Schechter fits of the radial bin LFs with green colour gradient (dark-light inside-out). The total normalized fit of A\,85 is plotted in blue and the field one by Blanton et al. (2005) with an orange line. The right-hand panel shows the three different aperture normalized fits with similar colour code: the green gradient inside-out, the total one in blue and the same field fit in orange. In both panels the grey dashed vertical lines indicates the magnitudes ranges for the bright ($M_r \leq -21.5$), intermediate ($-21.5 < M_r < -18$) and dwarf ($M_r \geq -18$) galaxies}
\label{lf_norm}
\end{figure*}

In order to understand the meaning of these variations, we studied the normalized LFs. The normalization factor used is the integral of each LF -- i.e. within $\left[ -22.5 \leq M_r \leq -16 \right]$. Fig. \ref{lf_norm} shows the normalized fitted LFs. For sake of clarity we plot only the fitted functions to highlight the differences.

With a visual inspection of the normalized LFs (see Fig. \ref{lf_norm}), we divided the galaxy population of A\,85 into three groups: bright ($M_r \leq -21.5$), intermediate ($-21.5 < M_r < -18$) and dwarf ($M_r \geq -18$) galaxies. The former and the latter subsets do not present significant dependence on the environment. Furthermore, they are invariant with respect to the field \citep[see LF by ][]{blanton2005}. These results are corroborated by the Kolmogolov--Smirnov test with a probability always larger than 10\% for the two galaxy groups. Therefore, the null hypothesis cannot be rejected, and we cannot exclude that they come from a single population. However, this is not true for the intermediate galaxies that show a clear environmental dependence. 

\subsection{LF of the different galaxy populations}
The morphology--density relation proposed by \cite{dressler1980} suggests a radial dependence of the LF. In this section, we first analyse the global LF quantifying the relative contribution of the galaxy subsets. We then look to the radial dependence for the same subsets.

\begin{figure*}
\centering
 \includegraphics[width=1\linewidth]{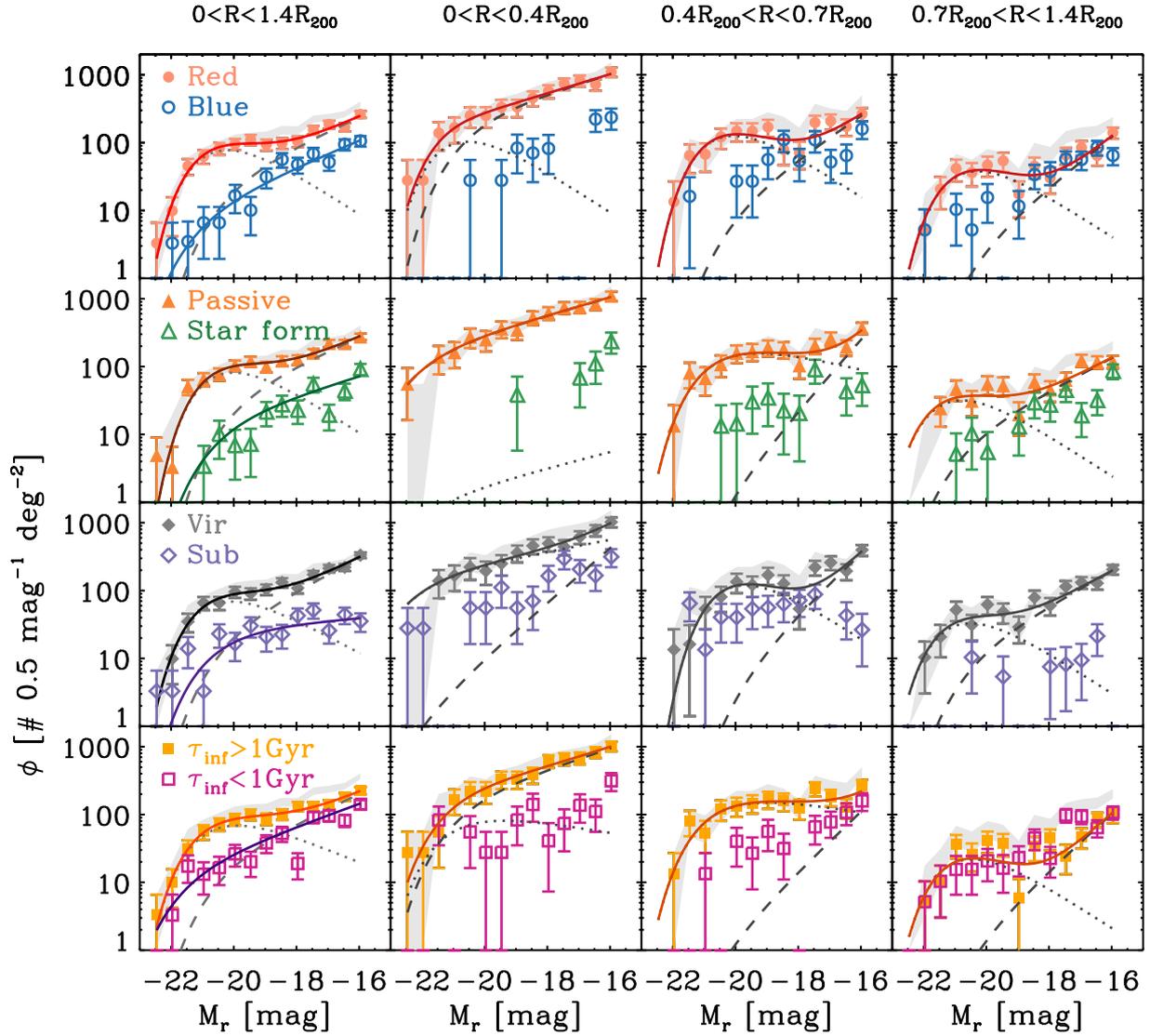}
 \caption{We show the total (first column) and radially binned LFs of the following populations: red and blue (first row), passive and star forming (second row), virialized and not virialized (third row), and early and late infall (forth row) galaxies. The LF values are plotted with their errors and the symbols and colours are as in Fig. \ref{ra_dec_pop}. The grey shadow is the value of the correspondent global LF. The solid lines are the fits, and the dotted and dashed lines are the bright and faint components of the double Schechter respectively (when it applies).}
\label{lf_pop_rad}
\end{figure*}

In the first column of Fig. \ref{lf_pop_rad}, we show the global LFs. The most abundant galaxies in the whole magnitude range are red, passive, virialized and early infall in the cluster history. These populations contribute to the 75 -80\% of the total luminosity density of the cluster (see table \ref{tab_pop_fit}). Their LFs present an upturn at the faint end similar to the one showed by the global LF and are modelled by a double Schechter function. Blue, star-forming, located in substructures and recent infall members contribute to only [15 - 25 \%] of the total luminosity density. Their LFs are well represented by a single Schechter function, and the values of the fit parameters are given in table \ref{tab_pop_fit}.

The radial LFs are shown in the other columns of Fig. \ref{lf_pop_rad} and the fit parameters are given in table \ref{tab_pop_rad_fit1}. Note that the fits are double Schechter functions and were performed only on the dominant populations due to the poor statistics of the other ones. Furthermore, we applied the same aperture analysis obtaining results similar to the global LFs. 

Focusing on Fig. \ref{lf_pop_rad}, the general trends of the red, passive, virialized and early infall LFs are the same of the global cluster. In particular, they present the two components with similar relative contributions, the upturn in the outskirts and the dip moving towards brighter magnitude from the intermediate to the outer region. Moreover, the external LFs present a mixture of the two subsets of galaxies, especially at the faint end. Consequently, the  upturn of the outer LF of A\,85 is due to both red and blue, passive and star-forming, and early and late infall galaxies. 

\begin{figure*}
\centering
 \includegraphics[width=1\linewidth]{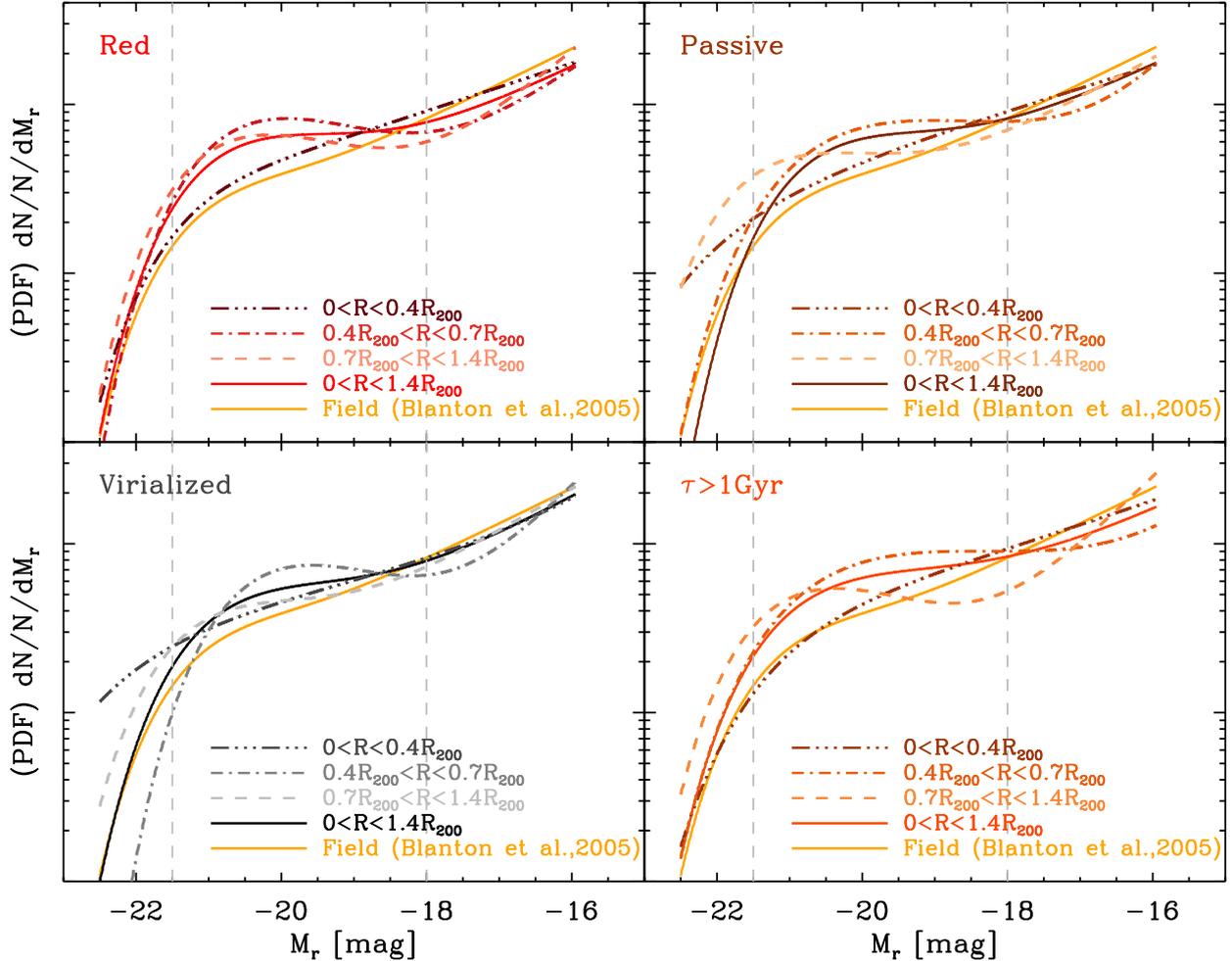}
 \caption{The plots show the radial bin normalized fits of the dominant populations in colour gradients (dark-light inside-out) and with different linestyles according to the legend, the total  and the field (global, by Blanton et al., 2005) normalized fits in solid lines.}
\label{lf_pop_norm}
\end{figure*}

In Fig. \ref{lf_pop_norm}, we show the radial normalized LFs of the dominant populations: red, passive, virialized and early infall. \textbf{The normalization factor is the integral of each LF. Fig. \ref{lf_pop_norm}} shows features like the global LF of A\,85. Indeed, we can identify the three groups of galaxies -- bright, intermediate and dwarfs -- and their similar trends with the environment (compare with Fig. \ref{lf_norm}).

\begin{table}
\begin{center}
\begin{tabular}{ccccc}
\hline
\hline
$M_{r}^* $ (mag) &$\alpha_\text{b} $ &$\alpha_\text{f}$  & $L_\text{b}/L_\text{t}$  &$L_\text{f}/L_\text{t}$\\
\hline
\multicolumn{5}{c}{$0 \, R_{200} < \,R\, \leq 1.4 \, R_{200}$}\\[5pt]
$-20.64 \; _{-0.08}^{+0.09}$ &$-0.31 \; _{-0.10}^{+0.11} $ &$ -1.47 \; _{-0.02}^{+0.02}$  &$71\% $  &$29\%$\\[5pt]
\hline
\multicolumn{5}{c}{$0 < \,R\,\leq 0.4 \, R_{200}$}\\[5pt]
$-21.36 \; _{-0.05}^{+0.05}$ &$-1.22 \; _{-0.01}^{+0.01} $ &$ -2.23 \; _{-0.03}^{+0.04}$  &$98\% $  &$2\%$\\[5pt]
\multicolumn{5}{c}{$0.4 \, R_{200} < \,R\, \leq 0.7 \, R_{200}$}\\[5pt]
$-20.71 \; _{-0.06}^{+0.07}$ &$-0.80 \; _{-0.02}^{+0.02} $ &$ -2.30 \; _{-0.03}^{+0.03}$  &$97\% $  &$3\%$\\[5pt]
\multicolumn{5}{c}{$0.7 \, R_{200} < \,R\, \leq 1.4 \, R_{200}$}\\[5pt]
$-20.77 \; _{-0.14}^{+0.14}$ &$-0.25 \; _{-0.12}^{+0.14} $ &$ -1.64 \; _{-0.02}^{+0.02}$  &$78\% $  &$22\%$\\[5pt]
\hline
\multicolumn{5}{c}{$0 < \,R\, \leq 0.7 \, R_{200}$}\\[5pt]
$-21.02 \; _{-0.06}^{+0.06}$ &$-1.03 \; _{-0.01}^{+0.01} $ &$ -2.25 \; _{-0.03}^{+0.03}$  &$97\% $  &$3\%$\\[5pt]
\hline
\end{tabular}
\caption{Schechter parameters of the total LF of A\,85, of the radial LFs and of the middle aperture in the last row (the smaller aperture is the same as the inner radial bin and the larger as the global). The last two columns are the percentage of the integrated luminosity of the bright and the faint population with respect to the total integrated luminosity. \label{tab_rad_fit}}
\end{center}
\end{table}

\begin{table}
\begin{center}
\begin{tabular}{cccc}
\hline\hline
$M_{r}^* $ [mag] &$\alpha_b $  &$\alpha_f$ &$L_i/L_t$\\
\hline
\multicolumn{4}{c}{Red}\\[5pt]
$-20.52\; _{-0.09}^{+0.10}$  &$-0.25\; _{-0.09}^{+0.11}$ &$-1.53 \; _{-0.02}^{+0.02} $ &87\% \\[5pt]
\multicolumn{4}{c}{Blue}\\[5pt]
$-21.38\; ^{+0.27}_{-0.28}$  &$-1.49\; ^{+0.04}_{-0.03}$ & -- &13\% \\[5pt]
\hline
\multicolumn{4}{c}{Passive}\\[5pt]
$-20.32\; _{-0.08}^{+0.09}$  &$-0.25\; _{-0.10}^{+0.11}$ &$-1.51 \; _{-0.02}^{+0.02} $ & 89\% \\[5pt]
\multicolumn{4}{c}{Star forming}\\[5pt]
$-20.77 \; _{-0.23}^{+0.24}$ &$-1.36 \; _{-0.03}^{+0.03} $ & -- &11\% \\[5pt]
\hline
\multicolumn{4}{c}{Virialized}\\[5pt]
$-20.59 \; _{-0.09}^{+0.09}$ &$-0.37 \; _{-0.08}^{+0.09} $ &$ -1.59 \; _{-0.02}^{+0.02}$  &86\% \\[5pt]
\multicolumn{4}{c}{In substructures}\\[5pt]
$-20.79 \; _{-0.23}^{+0.28}$ &$-1.10 \; _{-0.03}^{+0.04} $ & -- &14\% \\[5pt]
\hline
\multicolumn{4}{c}{$\tau_\text{inf} > 1$Gyr}\\[5pt]
$-20.70\; _{-0.09}^{+0.10}$  &$-0.52\; _{-0.07}^{+0.08}$ &$-1.54 \; _{-0.02}^{+0.02} $ &75\% \\[5pt]
\multicolumn{4}{c}{$\tau_\text{inf} < 1$Gyr}\\[5pt]
$-21.83\; ^{+0.25}_{-0.25}$  &$-1.42\; ^{+0.02}_{-0.02}$ & -- &25\% \\[5pt]
\hline
\end{tabular}
\caption{Schechter parameters for the different galaxy populations. The last column is the percentage of the integrated luminosity of that population with respect to the total integrated one. Where only $\alpha_\text{b}$ is present, it means that the fit is a single Schechter, otherwise is a double Schechter function. \label{tab_pop_fit}}
\end{center}
\end{table}

\section{Discussion}
\subsection{LF components in A\,85}\label{lf_comp}
\cite{durret1999} studied the photometric LF of A\,85 and compared it to the LF of other nearby clusters. A dip in the LFs was found at similar magnitudes in these clusters. They concluded then that this similarity is due to the same composition of morphological types of galaxies in all clusters. Our LF of A\,85 shows that the dip has a radial dependence: no dip is observed in the inner LF, while in the intermediate and in the outer radial bins is located at $M_{r}=-18.0$ and $-$19.0, respectively.

The dip has similar radial trends in the LFs of red, passive, virialized and early-infall galaxies. It is clearly visible in the cluster outskirts, while it is not present in the inner radial bin (compare second, third and fourth columns of Fig. \ref{lf_pop_rad}). The relative contribution of the galaxy subsets changes with clustercentric distance. Blue, star-forming and recent arrival galaxy densities are smaller in the central region and similar in the external one. This feature indicates that the presence of the dip  can be the result of the evolution of galaxies in the cluster environment. Indeed, it is not present where a small fraction of galaxies is still star forming or blue. In other words, cluster members in the outer region have not yet fully experienced the effects of cluster environment. 

\cite{peng2010} studied the evolution of the mass function with redshift and the effects of the environment on it. They analysed different sets of galaxies and, in particular, they traced the blue and star-forming galaxies and the red and passive ones. They proposed a model where two types of quenching occur and depend on the mass of the galaxy itself. One of them affects massive galaxies and is called \textit{mass quenching}, while the other is effective at low-mass ranges and it is called \textit{environmental quenching}. In the case of passive and red objects, they needed a double Schechter function to model the data and concluded that the two mathematical components of the fit shape the two processes. Applying this interpretation to A\,85, we have two components that depend on the clustercentric distance and we called bright and faint components. We have the faint one with increasing weight towards the outskirts. Its relative contribution to the total luminosity in the radial bin (whole cluster) changes from the 2\% in the inner radial bin to the 22\% in the outer one. Similar trend is observed with respect to the total luminosity of the cluster, even if weakened: the faint population passes from 1 to 7\%. Consequently, the bright component shows a decreasing teen inside-out. The radial variation of the $M^*$ parameter highlight these different contributions. In particular, the trends of the red, blue, passive and star-forming subsets agree with the results of \cite{peng2010}. Moreover, the LFs of recent-arrival and not-virialized populations are fitted by a single Schechter, while the LFs of the respective dominant populations -- e.g. virialized and early-infall -- are better represented by a double Schechter. This suggests that the latter population has spent enough time within the cluster potential to have produced their two components. 

\subsection{Normalized LFs}
Normalizing the LF is useful in order to compare its shape among systems of different richness. A variety of normalization factors has been used in literature, and we analyse here their effects on our study.

We first normalized the LF by scaling it with total luminosity integrated over the full range of magnitudes. We note that it is common practice to restrict the integration over few magnitudes at the bright end of the LF. This choice is supported by the environmental invariance of the bright end \citep[][among others]{andreon1998,popesso2006}.

\begin{figure*}
\centering
\includegraphics[width=1\linewidth]{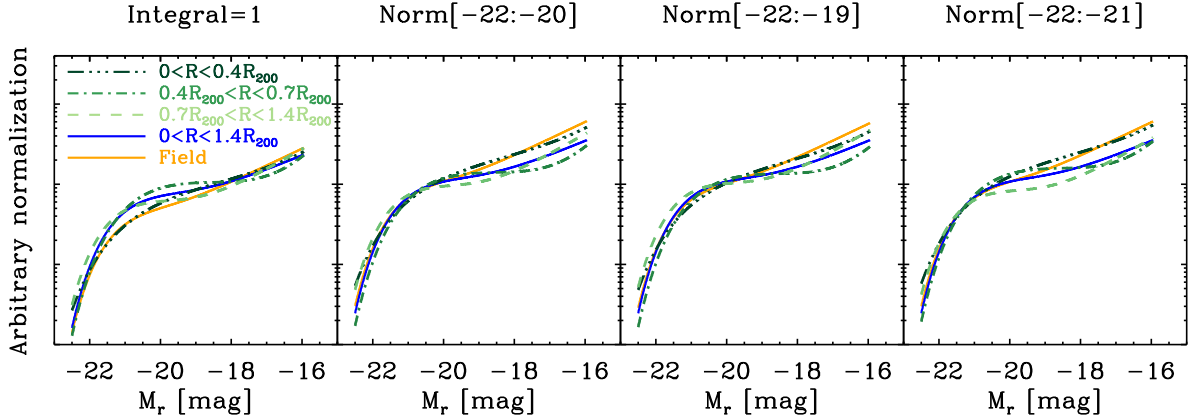}
\caption{Different normalizations are presented in this plot. From left to right: the one of this study, where the integral of the LF is one; normalization between $-$22 and $-$20; between $-$22 and $-$19; and between $-$22 and $-$21. The colours and the line styles are the same as the previous plots.}
\label{lf_cfr_norm}
\end{figure*}

We normalized the radial LFs of A\,85 using factors from the literature. In particular, we used the ranges $-22 < M_r < -20 $, $-22 < M_r < -19$, and $-22 < M_r < -21$ \citep[e.g.,][]{yagi2002,barkhouse2007}. As Fig. \ref{lf_cfr_norm} shows, we obtained similar faint-end slopes, but different relative number densities of dwarfs. Note that if the magnitude range used for the normalization includes intermediate luminosity galaxies, they are equalized by construction, so no studies on their radial properties can be made. Therefore, we strongly suggest to use the total luminosity as normalization factor and perform a complete and unbiased analysis of the environmental dependence of the LF. 

\subsection{The bright end of the LF}
The results on the bright end of the normalized LF of A\,85 confirm previous outcomes from the literature: it is independent of environment \citep[see e.g.][]{andreon1998,popesso2006,barkhouse2007}. This independence is likely a result of dynamical friction. Massive/luminous galaxies suffer of large dynamical friction entering the cluster potential, falling into the central region of the cluster on short time-scales. There, they merge with the BCG, giving a small fraction of luminous members ($M_r \sim -21.0$) in the inner radial bin. Note that dynamical friction can also explain other properties of this galaxy population observed and analysed in the literature \citep[e.g.][]{goto2005}.

\subsection{Environmental influence on dwarfs}
Dwarf galaxies are the most affected by environmental processes. Their shallow potential well impedes them from retaining their gas once in the cluster. For the same reason, tidal stripping efficiently remove stellar mass. The main physical processes acting on the gas component are starvation \citep{larson1980} and ram pressure \citep{gunn1972}, while the change in mass and/or morphology are mostly ascribed to harassment \citep{moore1996} and strong tidal interactions \citep{aguerri2009}. What the dominant processes in clusters are is still an open question, and the analysis of the faint end of the LF helps to put some constrains.

Several literature outcomes show a dependence on the environment of the faint-end slope on photometric LFs. Indeed, early-type LFs are steeper in the outer regions of a cluster \citep[see][]{popesso2006, barkhouse2009}. This result was recently confirmed with the spectroscopic stellar mass function of a medium redshift cluster \citep[see][]{annunziatella2016}. Photometric cluster LFs also show steeper slopes than the field \citep[see e.g.][]{popesso2006} indicating an excess of dwarf galaxies in high-density environments. This result is considered a probe supporting the evolutionary link between dwarf galaxies and clusters.

There are only a handful of deep, spectroscopic observations of nearby clusters, in which cluster-to-cluster variations of the shape of the LF are observed: Virgo and Abell 2199 LFs do not show any upturn at the faint end \citep{rines2008}, unlike A\,85 \citep{agulli2014}. Moreover, the radial analysis of \cite{rines2008} does not present a significant dependence of the LF on clustercentric radius, but opposite trends in the radial variation of $M^*$ between Virgo and Abell 2199. Indeed, the characteristic luminosity seems to be fainter at small radii for Abell 2199, while Virgo shows a trend similar to A\,85 (see Section \ref{lf_comp}). These observed cluster-to-cluster variations may depend on the formation history, the evolution of the large-scale environment or on the cluster mass. Indeed, \citet[][see in particular their Fig.1]{tully2002} hint a relation between the shape of the LF and the mass: more massive clusters show steeper slopes and more complex shapes. Being  A\,85 the most massive and its LF the steepest among the three clusters, this seems to indicate that we spectroscopically observed this relation. However, a wide spectroscopic survey of nearby clusters spanning a large range in masses will be necessary in order to put constrains on this and on the physical processes that give rise to these variations.

\cite{agulli2014} also studied the red and blue populations and compared their LFs to the field spectroscopic ones by \cite{blanton2005}. The two subsets show common features, including compatible slopes and the upturn for the red LFs. However, the dominant population at the faint end is different in the two environments: blue dwarfs are dominant in the field while red low-mass galaxies are dominant in the cluster. They concluded that the processes quenching star formation are playing a key role in this cluster. 

The present analysis on the LF of A\,85 finds similar results: red and passive LFs show the upturn, and the presence of the dip requires the modelling by a double Schechter function, while blue and star-forming galaxies are well represented by a single Schechter function. However, they have compatible faint-end slopes, suggesting again that the dominant mechanisms do not act on the size and/or the mass of the dwarfs. Indeed, a recent orbit analysis on simulated Virgo-like cluster found that only a small fraction (20-25\%) of low-mass galaxies suffers from substantial stellar mass loss due to harassment \citep{smith2015}. Moreover, our radial study found that the number density of late-type dwarfs is barely changing with radius, while the density of early-type galaxies is decreasing inside-out. These results indicate a transformation of the nature of dwarf galaxies due to the quenching of their star formation. In addition, the dominant mechanisms are more effective in the centre of the cluster where the hot intracluster medium is denser. The evolution of dwarfs is then led by starvation and ram-pressure stripping. 

Photometric studies of the LF in radial bins indicate a variation of the faint end for the red population \citep{popesso2006,barkhouse2009}. However, our analysis does not confirm this result. The normalized LFs shown in Fig. \ref{lf_pop_norm} highlight that the variation of the LF is at intermediate luminosities, whereas it is statistically insignificant at the faint end. The result that finds spectroscopical confirmation is the variation of $M^*$. Indeed, the characteristic luminosity of the red LF moves towards brighter values in the external cluster region (compare the top left panel of Fig. \ref{lf_pop_norm} with Fig. 11 of \cite{popesso2006}). However, due to cluster-to-cluster variations, a wider spectroscopic sample of nearby clusters is needed to quantitatively discuss the comparison between photometric and spectroscopic results.

\subsection{Intermediate galaxies}\label{ss_interm}
The results presented in this work show that the LF of A\,85 changes with the clustercentric distance at the intermediate magnitude range $(-21.5 \leq M_{r} \leq -18.0)$. In addition, cluster and field LFs differ within the same range, giving a clear environmental dependence for the evolution of intermediate luminosity galaxies. This dependence could be the consequence of the orbital distribution of these galaxies due to dynamical friction. Moreover, the large velocity dispersion of A\,85 inhibits mergers. Therefore, this type of galaxies will survive for a long time in eccentric orbits -- i.e. with large apocentres \citep[see ][]{biviano2004}.

The LF of A\,85 is dominated by red, passive and early-infall galaxies also at the magnitude range analysed here. These galaxies have passed near the cluster centre at least once \citep[see e.g.][]{muriel2014}, indicating that the intermediate galaxy population has suffered quenching of star formation due to the long time spent within the cluster potential. 

\section{Conclusions}
We have identified 460 cluster members of the nearby ($z$=0.055) cluster A\,85 down to $M > M^*+6$, thanks to deep spectroscopic observations. This data set allowed us to study the total and radial spectroscopic LFs of this cluster. The global LF and the LF dominant galaxy population -- i.e. red, passive, virialized and early-infall galaxies -- were fitted by a double Schechter function and they present a dip. These results point towards the presence of two galaxy populations in the cluster: a bright and a faint one. Their fraction depends on the clustercentric distance. Indeed, the faint component is more important in the outskirts of the cluster. In the central region of A\,85, the LFs do not show a dip and the faint galaxy populations only represent $\approx 2\%$ of the total galaxy density. Also blue and star-forming galaxies are a small fraction in this region. This result can indicate that the quenching of the star formation fills the dip observed in the external LFs. 

The normalized LFs of A\,85 show that the bright and faint ends are independent of the environment. Moreover, they are similar to the field ones. In contrast, the intermediate luminosity galaxies ($-21.5 \leq M_{r} \leq -18.0$) present an important dependences on the environment. 

Considering these results together, we suggest that hydrodynamical mechanisms are the main drivers of the dwarf evolution in A\,85. These processes transform blue low-mass galaxies into red ones with no significant changes in their mass/luminosity or size. The distribution of the bright and intermediate luminosity galaxies is due to dynamical friction. Therefore, the former suffer the largest dynamical friction, fall to the central region and eventually merge with the BCG on a short time-scale. Moreover, the bright end is independent of the environment. In contrast, the intermediate luminosity galaxies present a clear environmental dependence. This difference suggests that they survive in the cluster for several Gyr on eccentric orbits around the centre thanks to small dynamical friction.

\section*{Acknowledgements}
We acknowledge the anonymous referee for the useful comments that helped us to improve the paper. 

This work has been partially funded by the MINECO (grant AYA2013-43188-P).

CDV acknowledges support from the Ministry of Economy and Competitiveness (MINECO) through grants AYA2013-46886 and AYA2014-58308.

CDV and RB acknowledge financial support from the Spanish Ministry of Economy and Competitiveness (MINECO) under the 2011 Severo Ochoa Program MINECO SEV-2011-0187.

IA and AD acknowledge partial support from the INFN grant InDark and the grant PRIN 2012 "Fisica Astroparticellare Teorica" of the Italian Ministry of University and Research.

This research has made use of the Sixth Data Release of SDSS, and of the NASA/IPAC Extragalactic Database which is operated by the Jet Propulsion Laboratory, California Institute of Technology, under contract with the National Aeronautics and Space Administration.

The WHT and its service programme are operated on the island of La Palma by the Isaac Newton Group in the Spanish Observatorio del Roque de los Muchachos of the Instituto de Astrof\'isica de Canarias.







\appendix

\section[]{Tables}
\begin{table}
\begin{center}
\begin{tabular}{ccc}
\hline\hline
$M_{r}^* $ (mag) &$\alpha_\text{b} $ &$\alpha_\text{f}$\\
\hline\hline
\multicolumn{3}{c}{Red} \\
\hline\hline
\multicolumn{3}{c}{$0 < \,R\, \leq 0.4 \, R_{200}$}\\[5pt]
$-20.85\; _{-0.04}^{+0.04}$  &$-0.25\; _{-0.10}^{+0.11}$ &$-1.35 \; _{-0.00}^{+0.00} $ \\[5pt]
\multicolumn{3}{c}{$0.4 \, R_{200} < \,R\, \leq 0.7 \, R_{200}$}\\[5pt]
$-20.37\; _{-0.07}^{+0.08}$  &$-0.25\; _{-0.06}^{+0.06}$ &$-1.78 \; _{-0.02}^{0.02}$\\[5pt]
\multicolumn{3}{c}{$0.7 \, R_{200} < \,R\, \leq 1.4 \, R_{200}$}\\[5pt]
$-20.59\; _{-0.14}^{+0.16}$  &$-0.25\; _{-0.10}^{+0.12}$ &$-1.91 \; _{-0.04}^{+0.04} $ \\[5pt]
\hline
\multicolumn{3}{c}{$0 < \,R\, \leq 0.7 \, R_{200}$}\\[5pt]
$-20.61\; _{-0.07}^{+0.06}$  &$-0.32\; _{-0.08}^{+0.09}$ &$-1.40 \; _{-0.01}^{0.01}$\\[5pt]
\hline\hline
\multicolumn{3}{c}{Passive}\\
\hline\hline
\multicolumn{3}{c}{$0 < \,R\, \leq 0.4 \, R_{200}$}\\[5pt]
$-22.48\; _{-0.05}^{+0.06}$  &$-1.30 \; _{-1.23}^{+0.22}$ &$-1.33 \; _{-0.00}^{+0.00} $  \\[5pt]
\multicolumn{3}{c}{$0.4 \, R_{200} < \,R\, \leq 0.7 \, R_{200}$}\\[5pt]
$-20.68\; _{-0.07}^{+0.07}$  &$-0.73 \; _{-0.02}^{+0.02}$  &$-2.30 \; _{-0.03}^{+0.03} $ \\[5pt]
\multicolumn{3}{c}{$0.7 \, R_{200} < \,R\, \leq 1.4 \, R_{200}$}\\[5pt]
$-21.10\; _{-0.36}^{+0.26}$  &$-0.25\; _{-0.10}^{+0.12}$ &$-1.60 \; _{-0.04}^{+0.04} $ \\[5pt]
\hline
\multicolumn{3}{c}{$0 < \,R\, \leq 0.7 \, R_{200}$}\\[5pt]
$-20.56\; _{-0.06}^{+0.06}$  &$-0.63\; _{-0.04}^{+0.04}$ &$-1.59 \; _{-0.02}^{0.02}$\\[5pt]
\hline\hline
\multicolumn{3}{c}{Virialized}\\
\hline\hline
\multicolumn{3}{c}{$0 < \,R\, \leq 0.4 \, R_{200}$}\\[5pt]
$-22.61 \; _{-0.14}^{+0.13}$  &$-1.22 \; _{-0.01}^{+0.01}$ &$-2.01 \; _{-0.02}^{+0.02} $  \\[5pt]
\multicolumn{3}{c}{$0.4 \, R_{200} < \,R\, \leq 0.7 \, R_{200}$}\\[5pt]
$-20.00 \; _{-0.07}^{+0.08}$  &$-0.25 \; _{-0.06}^{+0.07}$ &$-2.09 \; _{-0.03}^{+0.03} $  \\[5pt]
\multicolumn{3}{c}{$0.7 \, R_{200} < \,R\, \leq 1.4 \, R_{200}$}\\[5pt]
$-20.82 \; _{-0.14}^{+0.15}$  &$-0.25 \; _{-0.12}^{+0.15}$ &$-1.63 \; _{-0.03}^{+0.02} $  \\[5pt]
\hline
\multicolumn{3}{c}{$0 < \,R\, \leq 0.7 \, R_{200}$}\\[5pt]
$-20.60\; _{-0.06}^{+0.06}$  &$-0.71\; _{-0.03}^{+0.03}$ &$-1.84 \; _{-0.02}^{0.02}$\\[5pt]
\hline\hline
\multicolumn{3}{c}{$\tau_{inf} > 1$Gyr}\\
\hline\hline
\multicolumn{3}{c}{$0 < \,R\, \leq 0.4 \, R_{200}$}\\[5pt]
$-21.17 \; _{-0.06}^{+0.06}$  &$-0.81 \; _{-0.11}^{+0.12}$ &$-1.39 \; _{-0.01}^{+0.01} $  \\[5pt]
\multicolumn{3}{c}{$0.4 \, R_{200} < \,R\, \leq 0.7 \, R_{200}$}\\[5pt]
$-20.76 \; _{-0.07}^{+0.07}$  &$-0.81 \; _{-0.02}^{+0.02}$ &$-2.09 \; _{-0.04}^{+0.05} $  \\[5pt]
\multicolumn{3}{c}{$0.7 \, R_{200} < \,R\, \leq 1.4 \, R_{200}$}\\[5pt]
$-20.77 \; _{-0.21}^{+0.21}$  &$-0.25 \; _{-0.12}^{+0.16}$ &$-2.04 \; _{-0.06}^{+0.05} $  \\[5pt]
\hline
\multicolumn{3}{c}{$0 < \,R\, \leq 0.7 \, R_{200}$}\\[5pt]
$-21.10\; _{-0.07}^{+0.07}$  &$-1.05\; _{-0.02}^{+0.02}$ &$-2.08 \; _{-0.05}^{0.07}$\\[5pt]
\hline\hline
\end{tabular}
\caption{Schechter parameters of the radial LFs and of the middle aperture (the smaller aperture is the same as the inner radial bin) LFs for the dominant populations.\label{tab_pop_rad_fit1}}
\end{center}
\end{table}


\bsp	
\label{lastpage}
\end{document}